\documentclass[twocolumn,showpacs,preprintnumbers,amsmath,amssymb]{revtex4}
\usepackage{amsmath,amsfonts,latexsym,amssymb,graphics,epsfig,subfigure,color,makeidx}
\usepackage{xcolor}

\begin{document}
\title{Localization of Vector Field on Pure Geometrical Thick Brane}
\author{Tao-Tao Sui \footnote{suitt14@lzu.edu.cn},
        Li Zhao\footnote{lizhao@lzu.edu.cn, Corresponding author}
}
\affiliation{Institute of Theoretical Physics, Lanzhou University, Lanzhou 730000, China}

\begin{abstract}
In this paper, we investigate the localization of a five-dimensional vector field on a pure geometrical thick brane. In previous work, it was shown that a free massless vector field cannot be localized on such thick brane. Hence we introduce the interaction between the vector field and the background scalar field. Two types of couplings are constructed as examples. We get a typical volcano potential for the first type of coupling and a finite square-well-like potential for the second one. Both of the two types of couplings ensure that the vector zero mode can be localized on the pure geometrical thick brane under some conditions.
\end{abstract}

\maketitle

In the last two decades, the theory of higher-dimensional spacetime has provided a new insight for solving some relevant physics problems (e.g., the gauge hierarchy problem, dark matter, and the cosmological constant problem)
\cite{ADD,rs,Lykken,Rubakov,Randjbar-Daemi,KehagiasPLB2004}. The most famous model is the Randall-Sundrum (RS) brane scenario \cite{rs}. However, in the original RS model, the brane is very ideal because its thickness is neglected.
In later work, by considering the minimum length scale of a brane, more natural thick branes generated by one or more background scalar fields have been investigated \cite{dewolfe,guo1106,gremm,CsakiNPB2000,AguilarMPLA2010,YXLiuJHEP2010}.
In the brane world scenario, it is assumed that  gravity is free to propagate in the bulk and the zero modes of  all matter fields (electromagnetic, Yang-Mills, fermions etc.) are confined to the 3-brane for the purpose of matching with the  present gravitational and particle experiments. The assumption leads to an important question that how to realize the localization of various bulk fields on a brane. It is well known that a free massless scalar field can be localized on the Randall-Sundrum (RS) brane or its generalized branes \cite{BajcPLB2000,Liu0708,Koroteev08,Flachi09,Liu0907,Liu1101}. For a fermion field, without introducing the scalar-fermion coupling  \cite{ThickBrane2,guo1408,zhao1102,LiuXu2014,ThickBrane1,ThickBrane3,Liu0803,KoleyCQG2005,Neupane} or fermion-gravity coupling \cite{LiLiu2017a}, it does not have normalizable zero modes in five-dimensional RS-like brane models.
While for the vector field, when only considering the coupling between the vector field and the background metric, the zero mode of the vector field can be localized on the de Sitter brane \cite{LiuyxJHEP2008}, AdS brane \cite{zhao1406}, Bloch Brane \cite{zhao1402}, and some pure geometrical Weyl  branes \cite{Cendejas0603,YXLiuJHEP2010,Yang12,Yang14}, but in some other pure geometrical Weyl  brane cases \cite{Liu0708,ThickBrane1,Barbosa-Cendejas2005}, the zero mode of the vector field is non-normalized and thus cannot be localized on the Weyl branes. In order to get the localized vector zero mode on this pure geometrical branes \cite{Liu0708,ThickBrane1,Barbosa-Cendejas2005}, we need a new localization mechanism by introducing the interaction between the vector field and Weyl scalar.

We start with a five-dimensional pure geometrical Weyl brane
which is based on the following action \cite{Liu0708,ThickBrane1,Barbosa-Cendejas2005}
\begin{equation}
S_5^{\texttt{W}} =\int_{M_5^{\texttt{W}}}\frac{d^5x\sqrt{-g}} {16\pi G_5}
e^{\frac{3}{2}\omega} \left[R+3\tilde{\xi}(\nabla\omega)^2 +
6U(\omega)\right], \label{action}
\end{equation}
where $M_5^{\texttt{W}}$ is a five-dimensional Weyl-integrable spacetime which is
specified by the pair $(g_{MN},\omega)$ with a five-dimensional metric $g_{MN}$ and a Weyl scalar $\omega$. The parameter $\tilde{\xi}$ is a coupling constant, and  $U(\omega)$ is a self-interaction potential of $\omega$. In the Weyl frame, the affine connections $\Gamma_{MN}^P$ are defined as
\begin{eqnarray}
\label{connections}
\Gamma_{MN}^P =\{_{MN}^{\;\;P}\}-\frac{1}{2} (\nabla_{M}\omega
\delta_N^P+\nabla_{N}\omega \delta_M^P-g_{MN}\nabla^{P}\omega) \nonumber
\end{eqnarray}
with $\{_{MN}^{\;\;P}\}$ the Christoffel symbols.
The Weyl rescaling
\begin{eqnarray}
\label{weylrescalings}
 g_{MN}&\rightarrow&\Omega^{2} g_{MN},\\
 \omega~~&\rightarrow&\omega+\ln\Omega^2,\\
 \tilde{\xi}~~ &\rightarrow& \frac{\tilde{\xi}}{(1+\partial_\omega\ln\Omega^2)^2},
\end{eqnarray}
will break the invariance of the action (\ref{action}), where $\Omega^2$ is a smooth function of $\omega$ on $M_5^W$.
In order to keep the invariance of the Weyl action, we set
\begin{eqnarray}
U(\omega)=\lambda e^{\omega}
\end{eqnarray}
with $\lambda$  a constant parameter.

In this paper, we consider the five-dimensional flat brane solution in the pure geometrical Weyl gravity (\ref{action}). The metric that respects four-dimensional Poincar$\acute{e}$ invariance is
\begin{equation}
ds_5^2=e^{2A(y)}\eta_{\mu\nu}dx^\mu dx^\nu + dy^2,\label{linee}
\end{equation}
where $e^{2A(y)}$ is the warp factor, and $y$ stands for the extra coordinate.

By making the conformal transformation $\hat{g}_{MN}= e^{\omega}g_{MN}$, the Weyl affine connection becomes the Christoffel symbol,  and the Weyl Ricci tensor turns into the Riemannian Ricci tensor \cite{ThickBrane1}. Thus the Weyl frame (\ref{action}) can be mapped into a Riemannian form, where five-dimensional gravity is coupled to a scalar field with a self-interaction potential $\hat{U}(\omega)=e^{-\omega}U(\omega)$:
\begin{equation}
S_5^{\texttt{R}}=\int_{M_5^{\texttt{R}}}\frac{d^5x\sqrt{-\hat{g}}}{16\pi G_5} \Big[\hat{R}+3{\xi}\left(\hat{\nabla}\omega\right)^2+6
\hat{U}(\omega)\Big],\label{confaction}
 \end{equation}
where $\xi=\tilde{\xi}-1$. It should be noted that the hatted magnitudes and operators refer to the Riemann frame.
After the conformal transformation, the line element in the Riemann frame reads
\begin{equation}
d\hat{s}_5^2=e^{2\sigma(y)}\eta_{\mu\nu}dx^\mu dx^\nu+e^{\omega(y)}dy^2,\label{conflinee}
\end{equation}
where $2\sigma=2A+\omega$.  In the case of $Z_2$-symmetric thick brane with infinite extra dimension ($-\infty<y<+\infty$), the solutions for the warp factor and the scalar field read \cite{ThickBrane1}
\begin{equation}
e^{2A(y)}=[\cosh(ay)]^b,  ~\omega(y)=kb\ln\cosh(ay).\label{configuration1}
\end{equation}
where the parameters $a$ and $b$ are given by
\begin{equation}
a=\sqrt{\frac{4+3k}{1+k}2\lambda}, ~~~~ b=\frac{2}{4+3k},
\end{equation}
and $\lambda$ and $k$ satisfy the following constraints:
\begin{equation}
\lambda>0, ~~~~ k<-4/3.
\end{equation}
Therefore, the parameter $b$ is negative, and the warp factor is concentrated near the origin $y=0$.

In Ref. \cite{Liu0708}, the authors investigated the localization of a free massless vector field on the pure geometrical brane. The action for such vector field is given by \cite{Liu0708}
\begin{eqnarray}
S_1 = - \frac{1}{4} \int d^5 x \sqrt{-g} g^{M N} g^{R S} F_{MR}F_{NS}, \label{actionVector}
\end{eqnarray}
where $F_{MN} = \partial_M A_N - \partial_N A_M$ is the field strength for the vector field $A_M$.
For convenience, one changes the metric \eqref{linee} into a conformal from:
 \begin{equation}
 ds^2_5=e^{2A(z)}(\eta_{\mu\nu}dx^{\mu}dx^{\nu}+dz^2). \label{linee2}
\end{equation}
The relation between the new coordinate $z$ and the physical one $y$ is $dz = e ^{-A(y)}dy$.
For the above brane configuration, the zero mode $\widetilde{\rho_{0}}$ in the conformal flat coordinates defined in (\ref{linee2}) was obtained \cite{Liu0708}:
\begin{eqnarray}
\widetilde{\rho_{0}}(z)\propto\frac{1}{(1+3\lambda z^2)^{1/4}},\label{zero modeqw}
\end{eqnarray}
where the parameter has been set to $k=-5/3$.
Unfortunately, the integration related to the vector zero mode in the KK reduction is not convergent, which means that the vector zero mode \eqref{zero modeqw} is non-normalized. This results in that the vector zero mode cannot be localized on the thick brane, just as the RS brane case.

In order to obtain a localized vector zero mode, we introduce the interaction between the vector field and the background scalar field $\omega$. The general action for the vector field coupled with the background scalar is
\begin{equation}
S_1 = - \frac{1}{4} \int d^5 x \sqrt{-g} G(\omega)g^{M N} g^{R S} F_{MR}F_{NS},
\label{Coupl Action}
\end{equation}
where $G(\omega)$ is a function of the background scalar field $\omega$. Considering the background geometry (\ref{linee2}), the equations of motion read as
 \begin{eqnarray}
 e^{A}G(\omega)\partial_{\nu}(\eta^{\mu\nu}F_{\mu4})=0,\\
 e^{A}G(\omega)\partial_{\nu}\Big(\eta^{\nu\rho}\eta^{\mu\lambda}F_{\rho\lambda}\Big)
 +\eta^{\mu\lambda}\partial_{z}\Big(e^{A}G(\omega)F_{4\lambda}\Big)=0.
 \label{EOMs}
  \end{eqnarray}
According to Ref.~\cite{Liu0708}, we suppose that the $A_{M}$ satisfy the $Z_2$ symmetry with respect to the extra dimension $z$, which requires that $A_4$ has no zero mode in the effective four-dimensional theory. Moreover, for being consistent with the gauge invariant ($\oint dz A_4=0$), we set $A_4=0$ by using the  gauge freedom. Under the KK decomposition of the vector field
\begin{equation}
A_{\mu}(x^{\nu},z)=\sum_{n}a_{\mu}^{n}(x^{\mu})\tilde{\rho}_{n}(z),
\label{KK decomposition}
\end{equation}
and the orthonormality conditions
\begin{equation}
\int^{\infty}_{-\infty}dz e^{A}G(\omega)\tilde{\rho}_{m}(z)\tilde{\rho}_{n}(z)=\delta_{mn},
\label{orth conditions}
\end{equation}
the action \eqref{Coupl Action} can be reduced into an effective four-dimensional form:
\begin{eqnarray}
S_1 \!= \! \sum_{n}\!\!\int \!\!\!d^4 x\!\!
  \left(\!-\frac{1}{4} \eta^{\mu\lambda} \eta^{\nu\rho}
                 f_{\mu\nu}^{(n)} f_{\lambda\rho}^{(n)}
        \!-\!\frac{1}{2}m_{n}^2 \eta^{\mu\nu}a_{\mu}^{(n)}a_{\nu}^{(n)}\!\right),
\label{actionVector3}
\end{eqnarray}
where $f^{(n)}_{\mu\nu} = \partial_\mu a_\nu^{(n)} - \partial_\nu a_\mu^{(n)}$ is the four-dimensional field strength, and  $m_{n}$ is the mass of the $n$-th vector KK mode.

It can be shown that the extra dimensional part $\tilde{\rho}_{n}(z)$ satisfies the following differential equation
\begin{equation}
-\partial_{z}\Big(e^{A}G(\omega)\partial_{z}\tilde{\rho}_{n}\Big)=m^{2}_{n}e^{A}G(\omega)\tilde{\rho}_{n}.
\label{differential equation}
\end{equation}
For convenience, we rewrite $G(\omega)$ as $G(\omega)=e^{g(\omega)}$ with $g(\omega)$ a function of $\omega$.  By defining $\rho_{n}=e^{B/2}\tilde{\rho}_{n}$ with $B=A+g(\omega)$, Eq. (\ref{differential equation}) can be rewritten as the following Schr$\ddot{o}$dinger-like equation:
\begin{eqnarray}
  \left[-\partial^2_z +V(z) \right]\rho_{n}(z)=m^2\rho_{n}(z),  \label{Schrodiner equation}
\end{eqnarray}
where the effective potential $V(z)$ reads
\begin{eqnarray}
 V(z)=\frac{1}{2}\partial^2_z B+\frac{1}{4}(\partial_z B)^2.  \label{effective potential}
\end{eqnarray}
In fact, the Schr$\ddot{\mathrm{o}}$dinger-like equation (\ref{Schrodiner equation}) can also be written as
\begin{eqnarray}
\mathcal{H}\rho_{n}=m_{n}^2\rho_{n}
\end{eqnarray}
with the Hamiltonian operator given by
\begin{eqnarray}
\mathcal{H}=Q^\dag Q=(\partial_z+\frac{1}{2}\partial_z B)(-\partial_z+\frac{1}{2}\partial_z B)
\end{eqnarray}
As the operator $\mathcal{H}$ is positive definite, there are no tachyonic vector modes with negative $m_{n}^2$.

For the zero mode $\rho_{0}$, it can be solved from the Schr$\ddot{\mathrm{o}}$dinger-like equation (\ref{Schrodiner equation}) by setting $m^2=0$:
\begin{eqnarray}
 \rho_{0} = c_0 e^{B/2} =c_0 e^{(A+g(\omega))/2},  \label{ZeroMode}
\end{eqnarray}
where $c_0$ is the normalization constant. We know from Eq.~(\ref{orth conditions}) that, in order to obtain the effective action of the four-dimensional massless vector field $a_{\mu}^{(0)}$, the zero mode needs to satisfy the following normalization condition:
\begin{equation}
\int^{\infty}_{-\infty}dz |{\rho}_{0}(z)|^2 =c_0^2 \int^{\infty}_{-\infty}dz  e^{A+g(\omega)} = 1,
\label{normalizationCondition2}
\end{equation}
which can be transformed into the physical coordinate:
\begin{equation}
c_0^2 \int^{\infty}_{-\infty}dy  e^{g(\omega)} = 1.
\label{normalizationCondition3}
\end{equation}
Now it is clear that without the non-minimal coupling function $G(\omega)=g(\omega)$, the vector zero mode cannot be localized on the pure geometrical brane~\cite{Liu0708}.
In order to obtain the normalized zero mode, we need to choose the specific expression of $g(\omega)$.
In the following, we will introduce a simple expression $g(\omega)=\tau\omega$ and then put forward a  specific form $g(\omega)=\theta e^{\frac{3}{10}\omega}+\frac{3}{10}\omega$.

$\mathbf{Case \; I}$:\;\; $g(\omega)=\tau\omega$

\begin{figure}[!htb]
\includegraphics[width=0.35\textwidth]{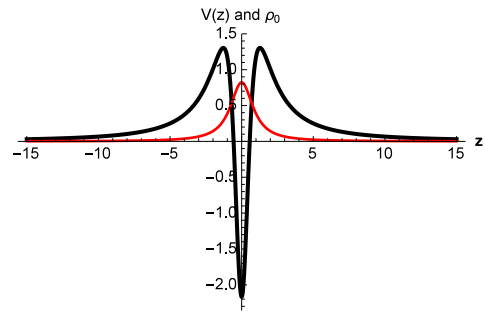}
\caption{The shapes of the effective potential $V(z)$ (thick black line ) of the vector KK modes
and the vector zero mode $\rho_{0}(z)$ (thin red line) for coupling I ($g(\omega)=\tau\omega$).
The parameters are set to $k=-5/3$, $\lambda=1/3$, and $\tau=-1$.}
\label{vector1}
\end{figure}

\begin{figure}[!htb]
\subfigure[$\lambda=1/3$]{
\includegraphics[width=0.35\textwidth]{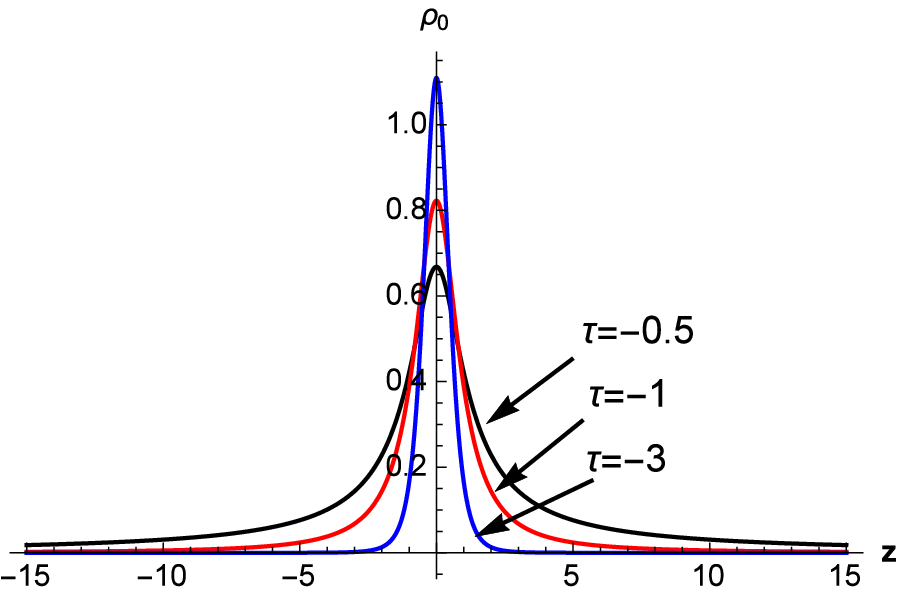}}
\subfigure[$\tau=-1$]{
\includegraphics[width=0.35\textwidth]{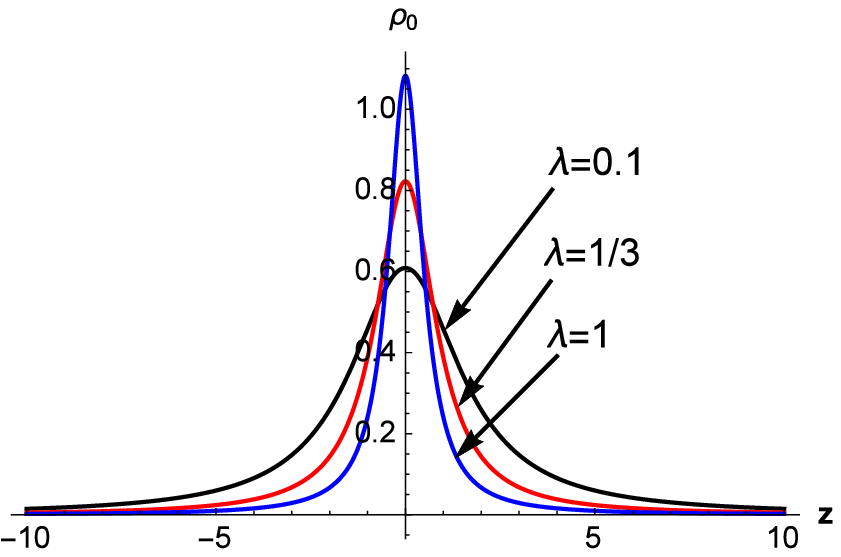}}
\caption{The effect of the parameters $\tau$ and $\lambda$ on the vector zero mode $\rho_{0}(z)$ for coupling I.
The parameter $k$ is set to $k=-5/3$.}
\label{vector11}
\end{figure}

We first assume the simple function $g(\omega)=\tau\omega$ with $\tau$ a constant. Such coupling between the background scalar filed and the vector field is called the dilaton coupling \cite{Chumbes1108,Cruz1211,Alencar1005,Alencar1008}. The effective potential $V(z)$ in (\ref{effective potential}) is
\begin{equation}
V(z)=\frac{\lambda(10\tau-3) \left[\lambda(10\tau-9) z^2+2\right]}{4 \left(3\lambda z^2+1\right)^2}.
\label{effective potential_I}
\end{equation}
It  has an asymptotic behavior that $V(z\rightarrow \pm\infty )\rightarrow 0$ and $V(z=0)=\frac{1}{2}\lambda (10\tau-3)$. The effective potential in this case is a typical volcano potential \cite{volcano,Davoudiasl}. It means that the effective potential provides a continuum gapless mass spectrum of the vector KK modes with positive $m^{2}_{n}>0$. The expression of the zero mode $\rho_{0}(z)$ is
\begin{equation}
\rho_{0}=c_0 \left(1+ 3\lambda z^2\right)^{\frac{5\tau}{6}-\frac{1}{4}}.
\label{zero mode1}
\end{equation}
The normalization condition \eqref{normalizationCondition2} requires
\begin{equation}
\tau<0.
\end{equation}
With the above condition, the integration in \eqref{normalizationCondition2} is convergent for infinite extra dimension.
The normalization constant is given by
\begin{equation}
c_0 =  \sqrt{\sqrt{\frac{3 \lambda }{\pi }}
             \frac{\Gamma \left(\frac{1}{2}-\frac{5 \tau }{3}\right)}
                  {\Gamma \left(-\frac{5 \tau}{3}\right)}
            }.
\end{equation}
Figure \ref{vector1} shows the shapes of the effective potential $V(z)$ and the vector zero mode $\rho_{0}(z)$ with a set of parameters $k=-5/3$, $\lambda=1/3$, and $\tau=-1$. Figure \ref{vector11} shows the effect of the parameters $\tau$ and $\lambda$ on the vector zero mode. It can be seen that, with the increase of $|\tau|$ or $\lambda$, the vector zero mode becomes higher but thinner and hence it is localized at a narrower region around the origin of the extra dimension $z=0$.

{$\mathbf{Case \; II}$:\;\;$g(\omega)=\theta e^{\frac{3}{10}\omega}+\frac{3}{10}\omega$}

\begin{figure}[!htb]
\includegraphics[width=0.35\textwidth]{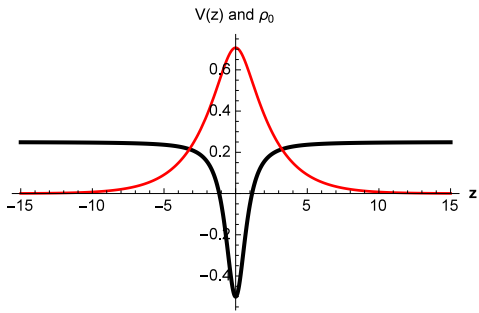}
\caption{The shapes of the effective potential $V(z)$ (thick black line)
and the vector zero mode $\rho_{0}(z)$ (thin red line) for coupling II. The
parameters are set to $k=-5/3$, $\lambda=1/3$, and $\theta=-1$.}
\label{vector2}
\end{figure}

\begin{figure}[!htb]
\subfigure[$\lambda=1/3$]{
\includegraphics[width=0.35\textwidth]{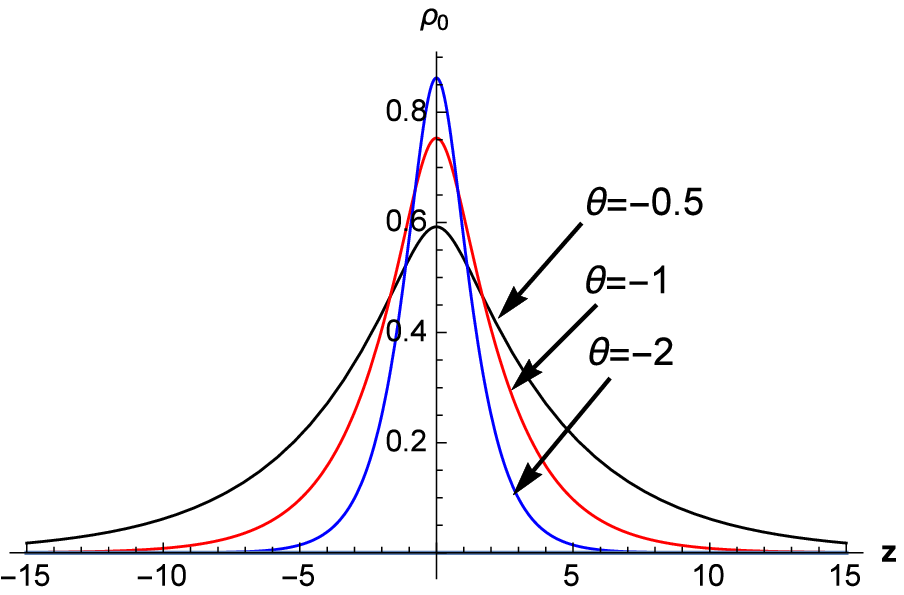}}
\subfigure[$\theta=-1$]{
\includegraphics[width=0.35\textwidth]{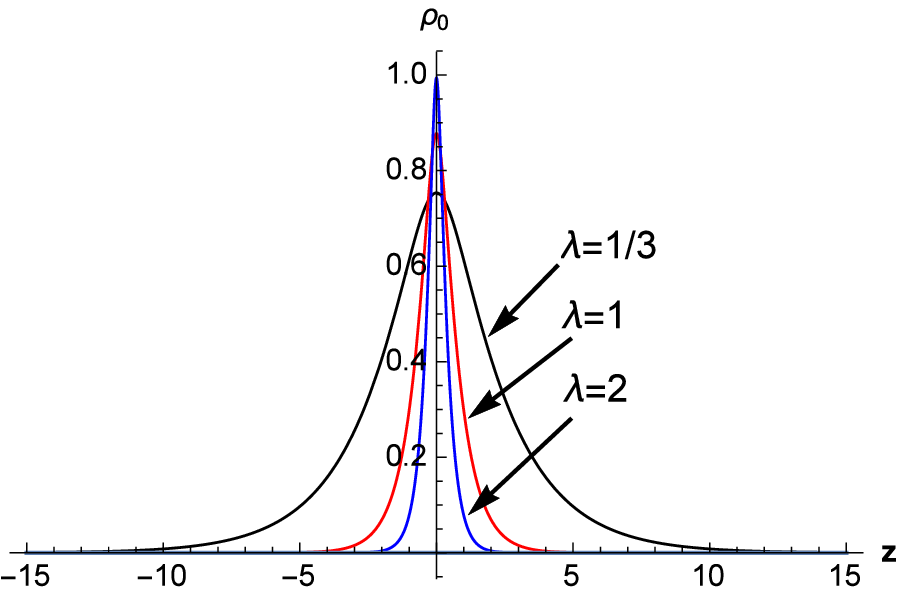}}
\caption{The effect of the parameters $\theta$ and $\lambda$ on the vector zero mode $\rho_{0}(z)$ for coupling II.
The parameter $k$ is set to $k=-5/3$.}
\label{vector21}
\end{figure}

Next, we consider the case of  $g(\omega)=\theta e^{\frac{3}{10}\omega}+\frac{3}{10}\omega$ with a dimensionless constant $\theta$. The expression of the effective potential $V(z)$ in (\ref{effective potential}) is now given by
\begin{equation}
V(z)=\frac{3\lambda \theta \left(3\lambda \theta z^2 \sqrt{3\lambda z^2+1}+2\right)}{4 \left(3\lambda z^2+1\right)^{3/2}},
\label{effectivepotential2}
\end{equation}
which has the asymptotic behavior of a finite square-well-like potential: $V(z\rightarrow \pm\infty )\rightarrow(\frac{3\lambda\theta}{2})^2$ and $V(z=0)=\frac{3\lambda\theta}{2}$. It means that the effective potential provides a mass gap to separate the zero mode from the KK modes. The expression of the zero mode $\rho_{0}(z)$  in this case reads
\begin{equation}
\rho_{0}(z)=c_{0}e^{\frac{1}{2} \theta \sqrt{3\lambda z^2+1}}.
\label{zero moed2}
\end{equation}
The integration in \eqref{normalizationCondition2} is convergent for infinite extra dimension if
\begin{equation}
\theta<0,
\end{equation}
and the normalization constant is
\begin{equation}
c_{0}=\sqrt{\frac{-\sqrt{3\lambda}~\theta}{2e^{{\theta}/{\sqrt{3\lambda}}}}}.
\end{equation}
The shapes of the effective potential $V(z)$ and the zero mode $\rho_{0}(z)$ are shown in Fig.~\ref{vector2} with the parameters $k=-5/3$, $\lambda=1/3$, and $\theta=-1$. Figure~\ref{vector21} gives a detailed description that the effect of the parameters $\theta$ and $\lambda$ on the vector zero mode. The vector zero mode has the similar phenomenon as case I that it becomes higher but thinner and is localized at a narrower region around $z=0$ with the increase of $|\theta|$ or $\lambda$.

To summarize, we have investigated the localization of a bulk vector field on a  pure geometrical flat thick brane. The interaction between the bulk vector field and the background scalar field, i.e.,  $-\frac{1}{4}e^{g(\omega)}F_{MN}F^{MN}$, was introduced. We  gave two forms of $g(\omega)$ to ensure that the zero mode of the vector field can be localized on the thick brane. For the first case $g(\omega)=\tau\omega$, we obtained a typical volcano potential. The localization conditoin is turned out to be $\tau<0$. The vector zero mode will be localized at a narrower region around $z=0$ with the increase of the coupling parameter $|\tau|$ or the model parameter $\lambda$. The effective potential for the second coupling function, i.e., $g(\omega)=\theta e^{\frac{3}{10}\omega}+\frac{3}{10}\omega$, has the asymptotic behavior of a finite square-well-like potential. The coupling parameter $\theta$ should be negative ($\theta<0$) in order to localize  the vector zero mode on the brane. The increasing $|\theta|$ or $\lambda$ can also make the vector zero mode close to a narrower region.

It is a great pleasure to thank Prof. Yuxiao Liu  for guidance and discussion. This work was supported by the National Natural Science Foundation of China (Grants Nos. 11522541 and 11375075) and the Fundamental Research Funds for the Central Universities
 (Grants No. lzujbky-2016-k04 and lzujbky-2014-31).

\end{document}